\documentstyle[aps,eqsecnum,preprint,floats,epsf,epsfig]{revtex}
\def\be{\begin{eqnarray}}
\def\en{\end{eqnarray}}

\def\non{\nonumber}
\def\la{\langle}
\def\ra{\rangle}
\def\ep{\varepsilon}
\def\ov{\overline}
\def\vma{{V-A}}

\def\Ova{O^q_{V-A}}

\def\Tva{T^q_{V-A}}

\def\lsim{ {\ \lower-1.2pt\vbox{\hbox{\rlap{$<$}\lower5pt\vbox{\hbox{$\sim$}
}}}\ } }

\begin{document}
\preprint{
\font\fortssbx=cmssbx10 scaled \magstep2
\hbox to \hsize{
\hfill$\raise .5cm\vtop{
                \hbox{}}$}
}
\draft
\vfill
\title{Phenomenological Analysis of $D$ Meson Lifetimes}
\draft
\vfill
\author{Hai-Yang Cheng and
Kwei-Chou Yang}
\address{Institute of Physics, Academia Sinica, Taipei, Taiwan 115, R.O.C.}

\maketitle
\date{May 1999}
\begin{abstract}
The QCD-based operator-product-expansion technique is
systematically applied to the study of charmed meson lifetimes. We
stress that it is crucial to take into account the momentum of the
spectator light quark of charmed mesons, otherwise the destructive
Pauli-interference effect in $D^+$ decays will lead to a negative
decay width for the $D^+$. We have applied the QCD sum rule
approach to estimate the hadronic matrix elements of color-singlet
and color-octet 4-quark operators relevant to nonleptonic
inclusive $D$ decays. The lifetime of $D_s^+$ is found to be
longer than that of $D^0$ because the latter receives a
constructive $W$-exchange contribution, whereas the hadronic
annihilation and leptonic contributions to the former are
compensated by the Pauli interference. We obtain the lifetime
ratio $\tau(D_s^+)/\tau(D^0)$ $\approx 1.08\pm 0.04$, which is
larger than some earlier theoretical estimates, but still smaller
than the recent measurements by CLEO and E791.
\end{abstract}

\vspace{0.7in}
\pacs{PACS numbers: 13.25.Hw, 12.38.Lg, 11.55.Hx, 12.39.Hg}
\section{Introduction}
It is well known that the observed lifetime difference between the
$D^+$ and $D^0$ is ascribed to the destructive interference in
$D^+$ decays and/or the constructive $W$-exchange contribution to
$D^0$ decays (for a review, see e.g., \cite{Bigi1}). By contrast,
the $D_s^+$ and $D^0$ lifetimes are theoretically expected to be
close to each other. For example, it is estimated in \cite{Bigi2}
that
\be
{\tau(D_s^+)\over\tau(D^0)}=\,1.00-1.07\,.
\en
However, the recent Fermilab E791 measurement of the $D_s^+$
lifetime yields $\tau(D_s^+)=0.518\pm 0.014\pm 0.007$ ps
\cite{E791}. When combining with the world average $D^0$ lifetime
\cite{PDG} yields the ratio
\be
{\tau(D_s^+)\over\tau(D^0)}=\,1.25\pm 0.04\, \qquad{\rm (E791)},
\en
which is different from unity by $6\sigma$. Meanwhile, the  CLEO
measurement of $D_s^+$ and $D^0$ lifetimes indicates
$\tau(D_s^+)=0.4863\pm 0.015\pm 0.005$ ps \cite{CLEO} and
\be
{\tau(D_s^+)\over\tau(D^0)}=\,1.19\pm 0.04\, \qquad{\rm (CLEO)},
\en
which is $5\sigma$ different from unity. Note that the $D^+_s$
lifetime measured by Fermilab and CLEO is better than the errors
of the world average value \cite{PDG} and that the lifetime ratio
of $D^+_s$ to $D^0$ is larger than the previous world average
\cite{PDG}:
\be
{\tau(D_s^+)\over\tau(D^0)}=\,1.13\pm 0.04\, \qquad{\rm (PDG)}.
\en

Based on the operator product expansion (OPE) approach for the
analysis of inclusive weak decays of heavy hadrons, it is known
that the $1/m_c^2$ corrections due to the nonperturbative kinetic
and chromomagnetic terms are small and essentially canceled out in
the lifetime ratios. By contrast, the $1/m_c^3$ corrections due to
4-quark operators can be quite significant because of the
phase-space enhancement by a factor of $16\pi^2$. The nonspectator
effects of order $1/m_c^3$ involve the Pauli interference in $D^+$
decay, the $W$-exchange in $D^0$ decay, the $W$-annihilation and
Cabibbo-suppressed Pauli interference in nonleptonic $D_s^+$.
While the semileptonic decay rates of $D^+,D^0$ and $D_s^+$ are
essentially the same, there is an additional purely leptonic decay
contribution to $D_s^+$, namely $D^+_s\to\tau\bar\nu$. The
dimension-6 four-quark operators which describe the nonspectator
effects in inclusive decays of heavy hadrons are well known
\cite{Bigi92,BS93}. However, it is also known that there is a
serious problem with the evaluation of the destructive Pauli
interference $\Gamma^{\rm int}(D^+)$ in $D^+$. A direct
calculation indicates that $\Gamma^{\rm int}(D^+)$ overcomes the
$c$ quark decay rate so that the resulting nonleptonic decay width
of $D^+$ becomes negative \cite{BS94,Chern}. This certainly does
not make sense. This implies that the $1/m_c$ expansion is not
well convergent and sensible, to say the least. In other words,
higher dimension terms are in principle also important. It has
been conjectured in \cite{BS94} that higher-dimension corrections
amount to replacing $m_c$ by $m_D$ in the expansion parameter
$f_D^2m_D/m_c^3$, so that it becomes $f_D^2/m_D^2$. As a
consequence, the destructive Pauli interference will be reduced by
a factor of $(m_c/m_D)^3$.

Another way of alleviating the problem is to realize that the
usual local four-quark operators are derived in the heavy quark
limit so that the effect of spectator light quarks can be
neglected. Since the charmed quark is not heavy enough, it is very
important, as stressed by Chernyak \cite{Chern}, for calculations
with charmed mesons to account for the nonzero momentum of
spectator quarks. It turns out that the Pauli interference in
$D^+$ decay is suppressed by a factor of $(\la p_c\ra-\la
p_d\ra)^2/\la p_c\ra^2=(\la p_D\ra-2\la p_d\ra)^2/m_c^2$, where
$\la p_c\ra$ and $\la p_d\ra$ are the momenta of the $c$ and $\bar
d$ quarks, respectively, in the $D^+$ meson. Because the charmed
quark is not heavy, the spectator $\bar d$ quark carries a sizable
fraction of the charmed meson momentum. Consequently, the Pauli
effect in $D^+$ decay is subject to a large suppression and will
not overcome the leading $c$ quark decay width. Based on this
observation, in the present paper we will follow \cite{Chern} to
take into account the effects of the spectator quark's momentum
consistently. In the framework of heavy quark expansion, this
spectator effect can be regarded as higher order $1/m_c$
corrections.

In order to understand the $D$-meson lifetime pattern, it is
important to have a reliable estimate of the hadronic matrix
elements. In the present paper we will employ the QCD sum rule to
evaluate the unknown hadronic parameters $B_1,B_2,\ep_1,\ep_2$, to
be introduced below.  In Sec.~\ref{sec:GF}, we will outline the
general framework for the study of the charmed meson lifetimes.
Then in Sec.~\ref{sec:sum rules} we proceed to compute the
hadronic parameters using the sum rule approach. Sec.~\ref{sec:DC}
presents results and discussions.

\section{General Framework}\label{sec:GF}
The inclusive nonleptonic and semileptonic decay
rates of a charmed meson to order $1/m_c^2$ are given by \cite{Bigi92,BS93}
\be
\label{nlspec} \Gamma_{\rm NL,spec}(D) &=& {G_F^2m_c^5\over
192\pi^3}N_c\,V_{\rm CKM}\, {1\over 2m_D} \Bigg\{
\left(c_1^2+c_2^2+{2c_1c_2\over N_c}\right)-
 \Big[\alpha I_0(x,0,0)\la D|\bar cc|D\ra   \non \\
&-& {1\over m_c^2}I_1(x,0,0) \la D|\bar cg_s\sigma \cdot G c|D\ra
\Big]   -{4\over m_c^2}\,{2c_1c_2\over N_c}\,I_2(x,0,0) \la D|\bar
cg_s\sigma\cdot G c|D\ra\Bigg\},
\en
where $\sigma\!\cdot\! G=\sigma_{\mu\nu}G^{\mu\nu}$,
$x=(m_s/m_c)^2$, $N_c$ is the number of colors, the parameter
$\alpha$ denotes QCD radiative corrections \cite{Bagan}, and
\be
\label{sl} \Gamma_{\rm SL}(D) &=& {G_F^2m_c^5\over
192\pi^3}|V_{cs}|^2\,{ \eta(x,x_\ell,0)\over 2m_D} \nonumber \\
&\times& \Big[ I_0(x,0,0)\la D|\bar cc|D\ra-{1\over
m_c^2}\,I_1(x,0,0) \la D|\bar cg_s\sigma\cdot G c|D\ra \Big]  \,,
\en where $\eta(x,x_\ell,0)$ with $x_\ell=(m_\ell/m_Q)^2$ is the
QCD radiative correction to the semileptonic decay rate and its
general analytic expression is given in \cite{Hokim}. In
Eqs.~(\ref{nlspec}) and (\ref{sl}), $I_{0,1,2}$ are phase-space
factors (see e.g. \cite{Cheng} for their explicit expressions),
and the factor $V_{\rm CKM}$ takes care of the relevant
Cabibbo-Koyashi-Maskawa (CKM) matrix elements. In
Eq.~(\ref{nlspec}) $c_1$ and $c_2$ are the Wilson coefficients in
the effective Hamiltonian.

The two-body matrix elements in Eqs.~(\ref{nlspec}) and (\ref{sl})
can be parameterized as
\begin{eqnarray}
  \frac{\langle D|\bar cc|D \rangle}{2m_D} &=& 1
    - \frac{K_D}{2m_c^2}+\frac{G_D}{2 m_c^2} + O(1/m_c^3) \,,
\nonumber\\
  \frac{\langle D|\bar c{1\over 2}g_s\sigma\cdot G c|D
  \rangle}{2m_D} &=& {G_D} + O(1/m_c) \,,
\end{eqnarray}
where
\begin{eqnarray}
   K_D &\equiv& -\frac{\langle D|\bar h^{(c)}_v\, (iD_\perp)^2
h^{(c)}_v |D \rangle}{2m_D}=-\lambda_1\,,\nonumber\\
G_D &\equiv&
 \frac{\langle D|\bar h^{(c)}_v\,{1\over 2}g_s\sigma\cdot G
h^{(c)}_v |D \rangle}{2m_D}=3\lambda_2 \,.
\end{eqnarray}
The nonperturbative parameter $\lambda_2$ is obtained from the
mass squared difference of the vector and pseudoscalar mesons:
\be
(\lambda_2)_D &=& {3\over 4}(m^2_{D^*}-m^2_D)=0.138\,{\rm GeV}^2,
\non \\ (\lambda_2)_{D_s} &=& {3\over
4}(m^2_{D_s^*}-m^2_{D_s})=0.147\,{\rm GeV}^2.
\en
As for the parameter $\lambda_1$, it is determined from the mass
relation \cite{Bigi2}
\be
(\lambda_1)_{D_s}-(\lambda_1)_D\cong {2m_bm_c\over m_b-m_c}\left[
\ov m_{B_s}-\ov m_B-(\ov m_{D_s}-\ov m_D)\right],
\en
where $\ov m_P={1\over 4}(m_P+3m_{P^*})$ denotes the spin-averaged
meson mass. For $m_b=5.05$ GeV and $m_c=1.65$ GeV, we obtain
$(\lambda_1)_{D_s}-(\lambda_1)_D=-0.067\,{\rm GeV}^2$.

\begin{figure}[ht]
\vspace{1cm}
    \leftline{\hspace{1.1cm} \epsfig{figure=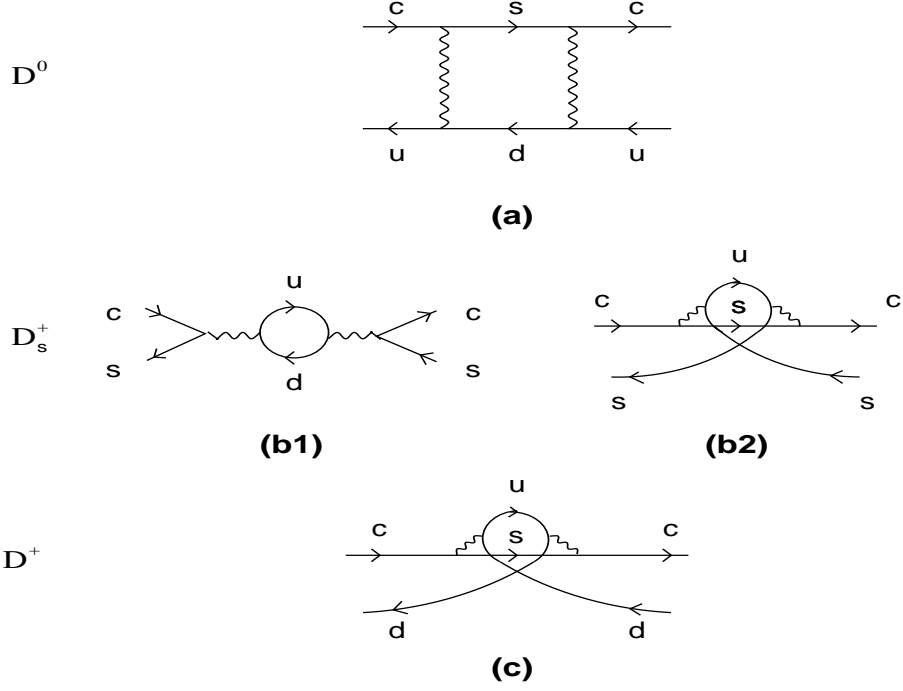,width=12cm,height=9cm}}
\vspace{0.9cm}
    \caption{Nonspectator effects: (a) $W$-exchange,
(b1) $W$-annihilation, (b2) and (c) Pauli interference.
\label{fig:nonspec}} \vspace{0.5cm}
\end{figure}

To the order of $1/m_c^3$, the nonspectator effects due to the
Pauli interference and $W$-exchange (see Fig. 1) may contribute
significantly to the lifetime ratios due to the two-body
phase-space enhancement by a factor of $16\pi^2$ relative to the
three-body phase space for heavy quark decay. As stressed in the
Introduction, it is crucial to invoke the effect of the light
quark's momentum in the charmed meson in order to properly
describe the $D$ lifetimes. For this purpose, the four-quark
operators relevant to inclusive nonleptonic $D$ decays are
\cite{Chern}
\be
\label{nsp} {\cal L}_{\rm NL,nspec} &=& {2G_F^2\over \pi}\,V_{\rm
CKM}\Bigg\{ g^{\mu\nu}k^2\eta_1 \left[ \left(2c_1c_2+{1\over
N_c}(c^2_1+c_2^2)\right)O_{\mu\nu}^d+2(c_1^2+ c_2^2)T_{\mu\nu}^d
\right]   \non \\ &+&{1\over 3}(k^\mu k^\nu \eta_2-k^2
g^{\mu\nu}\eta_3)\Big[ N_c \Big( c_2+{1\over N_c}c_1\Big)^2
O_{\mu\nu}^u+2c_1^2T_{\mu\nu}^u \non\\ &+& N_c\Big( c_1+{1\over
N_c}c_2\Big)^2 O_{\mu\nu}^s+2c_2^2T_{\mu\nu}^s \Big]\Bigg\},
\en
where
\be
O_{\mu\nu}^q &=& \bar c_L\gamma_\mu q_L\,\bar q_L\gamma_\nu c_L,
\non
\\ T_{\mu\nu}^q &=& \bar c_L\gamma_\mu t^a q_L\,\bar
q_L\gamma_\nu t^a c_L,
\en
with $t^a=\lambda^a/2$ and $\lambda^a$ being the Gell-Mann
matrices, and $\eta_1,~\eta_2,~\eta_3$ are phase-space factors,
depending on the number of strange quarks inside the loop of Fig.
1 \cite{Chern,NS}:
\be
(i)&& \qquad \eta_1=(1-x)^2,\qquad \eta_2=(1-x)^2(1+{x\over 2}),
\qquad \eta_3=(1-x)^2(1+2x), \non \\ (ii)&& \qquad \eta_1=(1-x)^2,
\qquad \eta_2=\sqrt{1-4x}\,(1-x), \qquad
\eta_3=\sqrt{1-4x}\,(1+2x),
\en
for (i) one strange quark and (ii) two strange quarks in the loop,
respectively, with $x=(m_s/m_c)^2$. Of course, $\eta_i=1$ in the
absence of strange loop quarks. In Eq.~(\ref{nsp}) the first term
proportional to $g^{\mu\nu}k^2$ contributes to the Pauli
interference, while the rest to the $W$-exchange or
$W$-annihilation, where $k$ is the total four-momentum of the
integrated quark pair \cite{Chern}. More specifically, $k=p_c+p_q$
for the $W$-exchange and $W$-annihilation, and $k=p_c-p_q$ for the
Pauli interference. In the heavy quark limit, $k\to p_c$ and it is
easily seen that (\ref{nsp}) is reduced to the more familiar form
\cite{NS}
\be
{\cal L}_{\rm NL,nspec} &=& {2G_F^2 m_c^2\over \pi}\,V_{\rm
CKM}\Bigg\{ \left(2c_1c_2+{1\over
N_c}(c^2_1+c_2^2)\right)\eta_1O_\vma^d+2(c_1^2+c_2^2)\eta_1
T_\vma^d \non
\\ &-& {1\over 3}N_c\Big( c_2+{1\over N_c}c_1\Big)^2( \eta_2O_\vma^u-\eta_3O_{
S-P}^u) -{2\over 3}c_1^2(\eta_2T_\vma^u-\eta_3T_{S-P}^u) \non\\
&-& {1\over 3} N_c\Big( c_1+{1\over N_c}c_2\Big)^2
(\eta_2O_\vma^s-\eta_3O_{ S-P}^s) -{2\over 3}
c_2^2(\eta_2T_\vma^s-\eta_3T_{ S-P}^s)\Bigg\},
\en
where use has been made of equations of motion, and
\begin{eqnarray}\label{4qops}
   O_{V-A}^q &=& \bar c_L\gamma_\mu q_L\,\bar q_L\gamma^\mu c_L
    \,, \nonumber\\
   O_{S-P}^q &=& \bar c_R\,q_L\,\bar q_L\,c_R \,, \nonumber\\
   T_{V-A}^q &=& \bar c_L\gamma_\mu t^a q_L\,
\bar q_L\gamma^\mu  t^a c_L \,, \nonumber\\
   T_{S-P}^q &=& \bar c_R\,t^a q_L\,\bar q_L\, t^ac_R \,,
\end{eqnarray}
with $q_{R,L}=(1\pm\gamma_5)q/2$.

In analog to the hadronic parameters defined in \cite{NS} for the
$B$ meson sector, we can also define four hadronic parameters
$B_1,B_2,\ep_1,\ep_2$ in the charm sector as
\be
\label{parameters1} {1\over 2m_{_{D_q}}}\la  D_q|\Ova| D_q\ra
&&\equiv {f^2_{D_q} m_{_{D_q}} \over 8}B_1\,, \nonumber\\ {1\over
2m_{_{D_q}}}\la D_q|\Tva| D_q\ra &&\equiv {f^2_{D_q}
m_{_{D_q}}\over 8}\varepsilon_1\,,
\en
and
\be
\label{parameters2} {k^\mu k^\nu\over 2m^3_{_{D_q}}} \la
D_q|O_{\mu\nu}^q| D_q\ra &&\equiv {f^2_{D_q} m_{_{D_q}}\over
8}B_2\,,\nonumber\\ {k^\mu k ^\nu\over 2m^3_{_{D_q}}} \la
D_q|T_{\mu\nu}^q| D_q\ra &&\equiv {f^2_{D_q} m_{_{D_q}}\over
8}\ep_2\,,
\en
for the matrix elements of these four-quark operators between $D$
meson states. Under the factorization approximation, $B_i=1$ and
$\varepsilon_i=0$ \cite{NS}.

The destructive Pauli interference in inclusive nonleptonic $D^+$
and $D_s^+$ decays and the $W$-exchange contribution to $D^0$ and
the $W$-annihilation contribution to $D^+_s$ are
\be
\label{bnonspec}
 \Gamma^{\rm exc}(D^0) = &-&\Gamma_0 \, \eta_{\rm
 nspec}\, (|V_{cs}|^2 |V_{ud}|^2+|V_{cd}|^2 |V_{us}|^2){m_D^2\over
 m_c^2}(1-x)^2\nonumber\\
 &&\times\Bigg\{ (1+{1 \over 2}x)\Big[({1\over
 N_c}c_1^2+2c_1c_2+N_cc_2^2)B_1+2c_1^2\ep_1 \Big]\non \\
 &&-(1+2x)\Big[({1\over N_c}c_1^2+2c_1c_2+N_cc_2^2)B_2
 +2c_1^2\ep_2\Big]  \Bigg\} \non\\
  &-&\Gamma_0 \, \eta_{\rm
 nspec}\, |V_{cs}|^2  |V_{us}|^2{m_D^2\over
 m_c^2}\sqrt{1-4x}\nonumber\\
 &&\times\Bigg\{ (1-x)\Big[({1\over
 N_c}c_1^2+2c_1c_2+N_cc_2^2)B_1+2c_1^2\ep_1 \Big]\non \\
 &&-(1+2x)\Big[({1\over N_c}c_1^2+2c_1c_2+N_cc_2^2)B_2
 +2c_1^2\ep_2\Big]  \Bigg\}\non\\
 &-&\Gamma_0 \, \eta_{\rm
 nspec}\, |V_{cd}|^2  |V_{ud}|^2{m_D^2\over
 m_c^2}\Bigg\{ ({1\over
  N_c}c_1^2+2c_1c_2+N_cc_2^2)(B_1-B_2)+2c_1^2(\ep_1-\ep_2)
 \Bigg\},\nonumber \\
 \Gamma^{\rm int}_-(D^+) &=&
 \Gamma_0\,\eta_{\rm nspec}|V_{ud}|^2 (|V_{cs}|^2(1-x)^2+|V_{cd}|^2)
 \,{(\la p_c\ra-\la p_d\ra)^2\over m_c^2}\non\\
 &&\times \left
 [(c_1^2+c_2^2)(B_1+6\ep_1)+6c_1c_2B_1\right],\nonumber\\
 \Gamma^{\rm ann}(D^+_s) &=& -\Gamma_0\eta_{\rm
 nspec} |V_{cs}|^2 |V_{ud}|^2\, \,{m_{D_s}^2\over m_c^2}\Bigg\{  ({1\over
 N_c}c_2^2+2c_1c_2+N_cc_1^2)(B_1-B_2)+2c_2^2(\ep_1-\ep_2) \Bigg\}
 \nonumber \\
 &&-\Gamma_0 \, \eta_{\rm
 nspec}\, |V_{cs}|^2|V_{us}|^2{m_{D_s}^2\over
 m_c^2}(1-x)^2\Bigg\{ (1+{1 \over 2}x)\Big[({1\over
 N_c}c_1^2+2c_1c_2+N_cc_2^2)B_1+2c_1^2\ep_1 \Big]\non \\
 &&-(1+2x)\Big[({1\over N_c}c_1^2+2c_1c_2+N_cc_2^2)B_2
 +2c_1^2\ep_2\Big]  \Bigg\} \,,\non\\
 \Gamma^{\rm int}_-(D^+_s)
 &=& \Gamma_0\,\eta_{\rm nspec}|V_{us}|^2
(|V_{cs}|^2(1-x)^2+|V_{cd}|^2)\,{(\la p_c\ra-\la
 p_s\ra)^2\over m_c^2} \non\\
 &&\times\left
 [(c_1^2+c_2^2)(B_1+6\ep_1)+6c_1c_2B_1\right],
\en
with
\be
\Gamma_0={G_F^2m_c^5\over 192\pi^3},~~~\eta_{\rm nspec}=16
\pi^2{f_{D_q}^2m_{D_q}\over m_c^3}.
\en

In Eq. (\ref{bnonspec}), $\la p_c\ra$ and $\la p_q\ra$ ($q=d,s$)
are the average momenta of the charmed and light quarks,
respectively, in the charmed meson. The sum $p_c+p_q$ can be
effectively substituted by $m_{D_q}$, the mass of the charmed
meson $D_q$. This can be nicely illustrated by the example of
$D_s\to\tau\bar\nu_\tau$ decay with the decay rate:
\begin{eqnarray}
 \Gamma(D_s\to \tau\bar \nu_\tau)
 \simeq
 \frac{G_F^2 m_\tau^2 f_{D_s}^2 m_{D_s}}{8\pi}|V_{cs}|^2
 \left( 1-\frac{m_\tau^2}{m_{D_s}^2}\right)^2 \,,
 \end{eqnarray}
an expression which can be found in the textbook. In the OPE
study, the same decay width is represented by
 \begin{eqnarray}
 \Gamma(D_s\to \tau\bar\nu_\tau)
 &&\simeq {G_F^2\over 6\pi} |V_{cs}|^2
 \left[ (p_c+p_{\bar s})^\mu (p_c+p_{\bar s})^\nu
 -g^{\mu\nu}(p_c+p_{\bar s})^2 +{3\over
 2}g^{\mu\nu}m_\tau^2
 \right]\nonumber\\
 &&\times{\langle D_s|(\bar c\gamma_\mu (1-\gamma_5)s) (\bar s\gamma_\nu
 (1-\gamma_5)c)|D_s\rangle \over 2m_{D_s}}
 \left( 1-\frac{m_\tau^2}{(p_c+p_{\bar s})^2}\right)^2 \,.
 \end{eqnarray}
Comparing the above two expressions, it is clear that
$(p_c+p_{\bar s})^2$ is nothing but $m_{D_s}^2$. Consequently,
$p_c-p_q$ can be approximated as $p_{D_q}-2p_q$ where $p_q$ could
be roughly set as the constituent quark mass $\sim 350$ MeV.
Compared to the naive OPE predictions, it is evident from Eq.
(\ref{bnonspec}) that the decay widths of $W$-exchange and
$W$-annihilation are enhanced by a factor of $(m_{D_q}/m_c)^2$,
whereas the Pauli interference is substantially suppressed by a
factor of $(p_{D_q}-2p_q)^2/m_c^2\sim 0.5\,$.

\section{QCD sum rule calculations of four-quark matrix
elements}\label{sec:sum rules} In order to calculate the
four-quark matrix elements appearing in the formula of the $D$
meson liftimes within the QCD sum rule approach, it is convenient
to adopt the following parametrization:
\be
&&\la D_q(p^D)|O^q_{\mu\nu}|D_q(p^D) \ra =(B p^D_\mu p^D_\nu +
\delta B\, g_{\mu\nu}m_{D_q}^2) {f_{D_q}^2\over
 4}\,,\nonumber\\
&&\la D_q(p^D)|T^q_{\mu\nu}|D_q(p^D) \ra  =(\ep p^D_\mu p^D_\nu +
\delta \ep\,  g_{\mu\nu} m_{D_q}^2)  {f_{D_q}^2\over4}\,,
\en
where the relations between $B, \delta B, \ep, \delta\ep$ and the
parameters $B_{1,2}, \ep_{1,2}$ defined in Eqs.
(\ref{parameters1}) and (\ref{parameters2}) are
\be
 &&B_1=B+4\delta B, \ \ B_2=B+\delta B\,,\nonumber\\
 &&\ep_1=\ep+4\delta \ep, \ \ \ep_2=\ep+\delta \ep\,.
\en

Unlike the $B$ meson case, the study of the $D$ meson is preferred
to begin with the full theory directly for several reasons: (1) In
the QCD sum rule study of the full theory, the working Borel
window of the $D$ meson case is about 2.0 GeV$^2<M^2< 3.0$
GeV$^2$. Hence, the extraction of relevant 4-quark matrix elements
can be obtained directly at the scale $\sim m_c$. (2) Since the
physical quantities expanded in $1/m_c$ will converge slowly due
to the fact that $m_c$ is not heavy enough, it becomes unnecessary
to work with the effective theory at the outset. (3) It is
customary in the literature to evolve the hadronic matrix elements
down to the confinement scale, say $\mu_h\sim 500$ MeV, in order
to apply the vacuum insertion hypothesis. However, as emphasized
in Ref.\cite{CY}, we shall avoid evaluating the matrix elements in
such a low scale because $\alpha(\mu_h)$ is of order unity at this
scale and large radiative corrections cannot be entirely grouped
into the Wilson coefficients.

We consider the following three-point correlation functions
\be\label{corr}
 &&\Pi^{O}_{\mu\nu}(p,p')=i^2\int dx\, dy\, e^{ipx-ip'y}
 \la 0|T\{[\bar q(x)i\gamma_5 c(x)]\, O^q_{\mu\nu}(0)\, [\bar
 q(y)i\gamma_5 c(y)]^\dagger\}|0\ra \,,\nonumber\\
 &&\Pi^{T}_{\mu\nu}(p,p')=i^2\int dx\, dy\, e^{ipx-ip'y}
 \la 0|T\{[\bar q(x)i\gamma_5 c(x)]\, T^q_{\mu\nu}(0)\, [\bar
 q(y)i\gamma_5 c(y)]^\dagger\}|0\ra \,.
\en
The sum rule calculation gives
  \be\label{had}
 &&{B p_\mu p'_\nu + \delta B\,  p\cdot p' g_{\mu\nu}\over
(p^2-m_{D_q}^2)(p'^2-m_{D_q}^2)}\biggl(\frac{f_{D_q}
m_{D_q}^2}{m_c+m_q}\biggr)^2  {f_{D_q}^2\over 4}\nonumber\\
 &&\simeq \frac{1}{4} p_\mu p'_\nu \Biggr\{
  {3\over 8\pi^2}\int^{s_0}_{m_c^2} ds
 \frac{1}{s-p^2} \bigg[ m_c\biggr(1-{m_c^2\over s}\biggl)^2+
m_q\biggr(1-{m_c^4\over s^2}\biggl)\bigg]  +{\langle \bar qq\rangle
\over p^2-m_c^2}\biggl(1+\frac{m_c m_q}{2(p^2-m_c^2)}\biggl)\nonumber\\
&&+{\langle g_s^2 G^2\rangle\over 48\pi^2 m_c}  \biggr( \frac{1}{p^2} -
\frac{1}{p^2-m_c^2} \biggl)
-\frac{m_c^2\langle g_s \bar q
\sigma  Gq\rangle}{2(p^2-m_c^2)^3}
\biggl(1+\frac{m_q}{4m_c}+\frac{m_c m_q}{8(p^2-m_c^2)} \biggr)
\Biggr\}^2\nonumber\\
&& + g_{\mu\nu} p\cdot p'\  \times{\cal O}({\rm dimension~8})\,,
\end{eqnarray}
and
\begin{eqnarray}
&&{\ep p_\mu p'_\nu + \delta \ep\,  p\cdot p' g_{\mu\nu}\over
(p^2-m_{D_q}^2)(p'^2-m_{D_q}^2)}\biggl(\frac{f_{D_q}
m_{D_q}^2}{m_c+m_q}\biggr)^2 {f_{D_q}^2\over 4}=-{1\over 3}
 (g_{\mu\nu} p\cdot p'- p_\mu p'_\nu)m_c^3\nonumber\\
&&\times \Biggl\{\frac{\la
 g_s^2 G^2 \ra}{(32\pi^2)^2} \biggl[ \int^{s_0}_{m_c^2}ds
 \int^{s_0}_{m_c^2}ds'
\frac{1}{(s-p^2)(s'-p'^2)}{1\over s^2 s'^2} [ m_c
(s+s'-m_c^2)-2m_q (2s+2s'-m_c^2)]\nonumber\\ &&
+2\frac{m_q}{m_c^2}\biggr( \frac{\ln [(m_c^2-p'^2)/(\mu
m_c)]}{p'^2-m_c^2} \int^{s_0}_{m_c^2}ds\frac{2s-3m_c^2}{s^2
(s-p^2)} + \frac{\ln [(m_c^2-p^2)/(\mu m_c)]}{p^2-m_c^2}
\int^{s_0}_{m_c^2}ds\frac{2s'-3m_c^2}{s'^2 (s'-p^2)} \biggr)
\biggr] \nonumber\\  && -\frac{\la g_s \bar q \sigma\cdot G q
\ra}{128\pi^2} \biggl[ \int^{s_0}_{m_c^2}ds
 \int^{s_0}_{m_c^2}ds'
\frac{1}{(s-p^2)(s'-p'^2)} \biggl({1\over s'^2}\delta (s-m_c^2) +{1\over
 s^2}\delta (s'-m_c^2) \biggr)\biggl(1-\frac{3m_q}{m_c} \biggr)\nonumber\\
&& -4\frac{m_q}{m_c^3}\biggr( \frac{\ln [(m_c^2-p'^2)/(\mu
m_c)]}{p'^2-m_c^2} \frac{1}{p^2-m_c^2} + \frac{\ln
[(m_c^2-p^2)/(\mu m_c)]}{p^2-m_c^2} \frac{1}{p'^2-m_c^2} \biggr)
\biggr] \Biggr\}\nonumber\\ &&+{\cal O} ({\rm dimension~6})\,,
\end{eqnarray}
where $\langle\cdots \rangle$
 stands for $\langle 0| \cdots |0\rangle$ and
\begin{eqnarray}
\langle 0|\bar qi\gamma_5 c|D_q\rangle =\frac{f_{D_q} m_{D_q}^2}{m_c+m_q}\,.
\end{eqnarray}
Here we have used the factorization (or vacuum insertion)
approximation to estimate the four-quark condensate. However,
since $\delta B$ does not receive four-quark operator
contributions under the factorization approximation, the
contribution from non-vacuum intermediate states may not be
negligible~\footnote{Up to dimension six, $\delta B$ is given by
 \be\label{NF}
 {\delta B\,  p\cdot p' \over
(p^2-m_{D_q}^2)(p'^2-m_{D_q}^2)}\biggl(\frac{f_{D_q}^2
 m_{D_q}^2}{m_c+m_q}\biggr)^2
 =\frac{m_c^2}{4}\langle \bar q\gamma_\mu(1-\gamma_5) q
 \bar q\gamma^\mu (1-\gamma_5)q\rangle
 \frac{1}{(p^2-m_c^2)(p'^2-m_c^2)},
 \en
which obviously vanishes under the factorization approximation. At
the confinement scale $\sim$500 MeV, the nonfactorizable
contribution due to the four-quark condensate was shown to be
sizable in \cite{Chern}. As a result, the lifetime ratio of
$\tau(D^+_s)/\tau(D^0)\sim 1.24$ obtained in \cite{Chern} is much
larger than previous estimates.}. We would like to remind readers
that the ratio of $\tau(D^+_s)/ \tau(D^0)$ is quite sensitive to
$\delta B$. We have estimated the four-gluon condensate
contribution to $\delta B$ and found that the enhancement of
$\delta B$ due to the four-gluon condensate is less $10^{-3}$ and
thus can be neglected. After performing the double Borel
transformations \cite{CY}, $p^2\to M^2$ and $p'^2\to M'^2$, on the
above sum rules and letting $M^2=M'^2$, we obtain

\begin{eqnarray}\label{rule1}
 B&&=4\biggl(\frac{m_c+m_q}{f_{D_q}^2 m_{D_q}^2}\biggr)^2
 e^{2m_{D_q}^2/M^2}\Biggr\{
  {3\over 8\pi^2}
\int^{s_0}_{m_c^2}ds e^{-s^2/M^2} \biggl[ m_c\biggl(1-{m_c^2\over
s}\biggr)^2 +m_q\biggl(1-{m_c^4\over s^2} \biggr)\biggr] \non \\
&& -\langle \bar qq\rangle \biggl(1-\frac{m_c
m_q}{2M^2}\biggl)e^{-m_c^2/M^2}  -{\langle g_s^2 G^2\rangle\over
48\pi^2 m_c}  \biggr( 1- e^{-m_c^2/M^2}\biggl)\nonumber\\
 &&+\frac{m_c^2\langle g_s \bar q \sigma\cdot
 Gq\rangle}{4M^4} \biggl( 1+{m_q\over 4m_c}-{m_c
m_q\over 24M^2}\biggr) e^{-m_c^2/M^2}\Biggr\}^2\,,\\ \delta B&&\approx 0\,,
\non \end{eqnarray}
and
\begin{eqnarray}\label{rule2}
  \ep &&=-\delta \ep
  =\biggl(\frac{m_c+m_q}{f_{D_q}^2 m_{D_q}^2}\biggr)^2
 {4\over 3}m_c^3 e^{2m_{D_q}^2/M^2} \int^{s_0}_{m_c^2}ds
 \int^{s_0}_{m_c^2}ds' e^{-(s+s')/M^2}\nonumber\\
&&\Biggl\{ \frac{\la
 g_s^2 G^2 \ra}{(32\pi^2)^2} {1\over s^2
 s'^2} \biggl[ m_c (s+s'-m_c^2)-2m_q (2s+2s'-m_c^2)\nonumber\\
&& +4m_q (2s-3m_c^2)s'\delta(s'-m_c^2) \biggl(\gamma+\ln\frac{(\mu
m_c)}{M^2}\biggr)\biggr]
 \nonumber\\
 && -\frac{\la g_s \bar q \sigma\cdot G q \ra}{128\pi^2}
\biggl[ \biggl({1\over
 s'^2}\delta (s-m_c^2) +{1\over
 s^2}\delta (s'-m_c^2) \biggr)\biggl(1-\frac{3m_q}{m_c} \biggr)\nonumber\\
&& +\frac{8m_q}{m_c^3}\delta(s-m_c^2)\delta(s'-m_c^2)
\biggl(\gamma+\ln\frac{(\mu m_c)}{M^2}\biggr) \biggr] \Biggr\} \,,
\end{eqnarray}
where $\gamma$ is the Euler's constant.

For numerical estimates of $B$ and $\ep$, we shall use the
following values of parameters:~\footnote{It is known that the
charmed quark mass used in the sum-rule studies is smaller than
the pole mass shown below. Likewise, the sum-rule decay constants
$f_D$ and $f_{D_s}$ are slightly smaller the values employed in
Sec. IV.} $f_{D_{u,d}}=170\pm 10$ MeV, $f_{D_s}=210\pm 10$ MeV,
$m_u=m_d=0$, $m_s=125\pm 25$~MeV, $m_c=1.40\pm 0.05$ GeV,
$s(D_{u,d})=6~{\rm GeV}^2$, $s(D_s)=6.5~{\rm GeV}^2$, and
\cite{CY}
\be
\label{condensate} &&\langle \bar uu\rangle_{\mu=1~{\rm GeV}}=
\langle \bar dd\rangle_{\mu=1~{\rm GeV}}=-(240\pm 20~{\rm
MeV})^3\,, \nonumber\\ &&\langle \bar ss\rangle= 0.8\times\langle
\bar uu\rangle\,,\nonumber\\
 &&\langle \alpha_s G^2 \rangle_{\mu=1~{\rm GeV}} =0.0377~{\rm
GeV^4} \,,\nonumber\\ &&\langle \bar qg_s\sigma\cdot G q\rangle=
(0.8~{\rm GeV^2})\times \langle \bar qq\rangle\,. \en Note that in
the sum rule study, $m_c$ is the current quark mass normalized at
$\mu^2=-m_c^2$.

To further improve the quality of the sum-rule results, we rescale
the nonperturbative quantities to the scale of the Borel mass $M$.
\begin{eqnarray}
 &&f_{D_q}(M)=f_{D_q}(m_c)\Bigl( {\alpha_s(M)\over
 \alpha_s(m_c)} \Bigr)^{-2/\beta_0} \,, \nonumber\\
 &&\langle \bar qq\rangle_{M} =\langle \bar qq\rangle_\mu \cdot \Bigl(
 {\alpha_s(M)\over \alpha_s (\mu)}
 \Bigr)^{-4/\beta_0}\,,\nonumber\\
 &&\langle g_s\bar q\sigma\cdot Gq\rangle_{M}=
 \langle g_s\bar q\sigma\cdot Gq\rangle_\mu \cdot
 \Bigl( {\alpha_s(M)\over \alpha_s(\mu)} \Bigr)^{2/(3\beta_0)}
 \,,\nonumber\\
 &&\langle \alpha_s G^2 \rangle_{M}= \langle
 \alpha_s G^2 \rangle_\mu\,,
 \end{eqnarray}
 where $\beta_0=\frac{11}{3}\,N_c-\frac{2}{3}\,n_f$ is the leading-order
 expression of the $\beta$-function with $n_f$ being the number of
 light quark flavors.

\begin{figure}[t]
\vspace{1cm}
    \leftline{\hspace{5cm}
\epsfig{figure=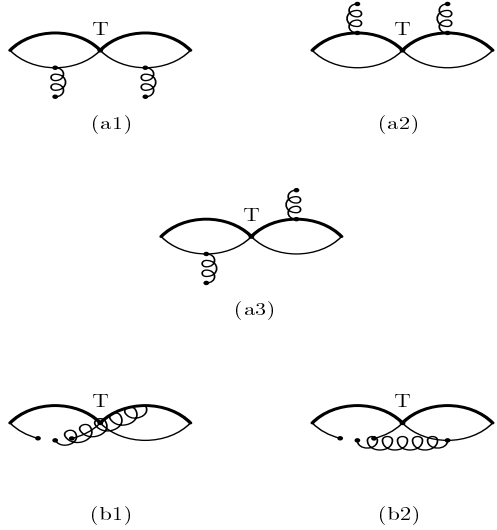,width=6.5cm,height=7cm}} \vspace{0.8cm}
    \caption{The main diagrams contributing to
the OPE series of ${\varepsilon}$ in Eq.~(\ref{rule2}): (a1)--(a3)
the gluon condensates, and (b1)--(b2) the quark-gluon mixed
condensates. The charmed quark is denoted by the heavy
line.\label{fig:OPE}} \vspace{0.5cm}
\end{figure}

\begin{figure}[t]
\vspace{1cm}
    \leftline{\hspace{2cm}
\epsfig{figure=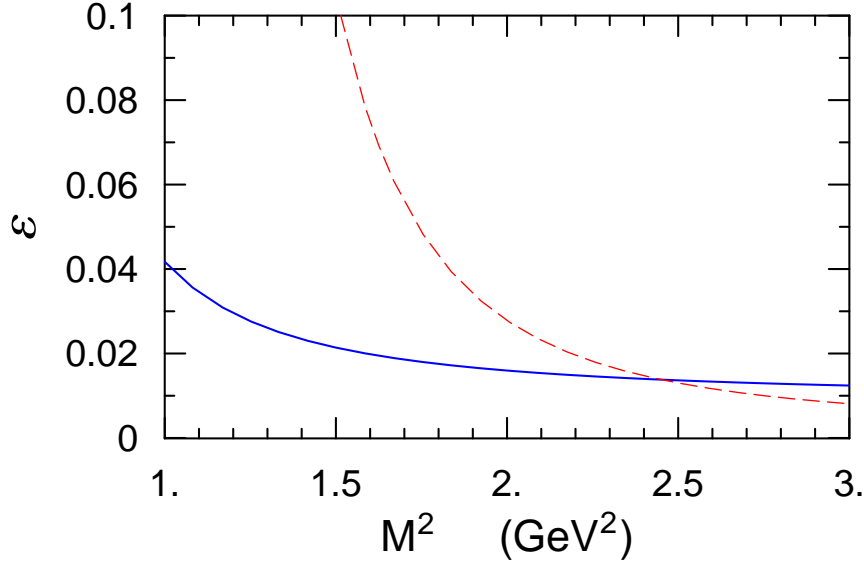,height=7.5cm}}
\vspace{0.7cm}
    \caption{$\varepsilon~(=-\delta \varepsilon)$ as a
function of the Borel mass squared $M^2$. The solid and dashed
curves are for $\varepsilon(D^{0,+})$ and $\varepsilon(D_s^+)$,
respectively. Here we have used $f_D=170$~MeV, $f_{D_s}=210$ MeV,
$m_c=1.40$~GeV, $m_s=125$~MeV, $s(D^{0,+})=6$ GeV$^2$,
$s(D_s)=6.5$ GeV$^2$, $\langle\bar qq\rangle_{\mu=1~{\rm
GeV}}=-(240~{\rm MeV})^3$, and
Eq.~(\ref{condensate}).\label{fig:epsilon}} \vspace{0.5cm}
\end{figure}

Let us explain the results obtained in Eq.~(\ref{rule1}) for the
parameter $B$ and Eq. (\ref{rule2}) for $\ep$. Eq.~(\ref{rule1})
can be approximately factorized as a product of two two-point
$f_{D_q}$ sum rules.  As a result, $B\approx 1$.  To the order of
dimension-five, the main contributions to the OPE series of
${\varepsilon}$ are depicted in Fig.~\ref{fig:OPE}, where we have
neglected the dimension-six four-quark condensate of the type
$\langle \bar q\Gamma\lambda^a q\ \bar q\Gamma\lambda^a q \rangle$
since its contribution is much less than that from dimension-five
or dimension-four condensates. The numerical result of
$\varepsilon$ ($=-\delta\varepsilon$) is shown in
Fig.~\ref{fig:epsilon}. Within the Borel window $2.0~{\rm
GeV}^2<M^2<3.0~{\rm GeV}^2$, we obtain $\varepsilon(D^{0,+})=$
$-\delta\varepsilon(D^{0,+})=0.015\pm 0.010$ and
$\varepsilon(D_s^+)=$
$-\delta\varepsilon(D_s^+)=0.015^{+0.015}_{-0.010}$, where the
error comes partially from the uncertainties of input parameters.
Consequently, $B_{1,2}$ and $\varepsilon_{1,2}$ are numerically
given by
\begin{eqnarray}
B_1=B_2\approx 1\,,\qquad\varepsilon_1(D^{0,+})=-0.045\pm 0.030\,,
\quad\varepsilon_1(D_s^+)=-0.045^{+ 0.045}_{-0.030}\,, \quad
\varepsilon_2=0\,. \label{para}
\end{eqnarray}
Since the sum rule calculation is built upon the quark-hadron
duality hypothesis, it is difficult to estimate the intrinsic
errors in this approach. However, if the OPE series is extended to
higher dimension operator terms, then the errors will be improved.
Moreover, it is desirable to evaluate the non-vacuum intermediate
state contributions to $\delta B$ =$3(B_1-B_2)$ as the ratio of
$\tau(D^+_s)/ \tau(D^0)$ is quite sensitive to $\delta B$. Even if
$\delta B$ deviates from zero by a small amount, say 0.005, the
ratio $\tau(D^+_s)/ \tau(D^0)$ will be enhanced by 6\%.

\section{Results and Discussions}\label{sec:DC}
The total decay width of the charmed meson is given by
\be
\Gamma(D)=\Gamma_{\rm NL,spec}+\Gamma_{\rm NL, nspec}+\Gamma_{\rm
SL}+\Gamma_{\rm lep},
\en
where $\Gamma_{\rm NL,spec}$ and $\Gamma_{\rm NL, nspec}$ denote
nonleptonic decay widths [cf. Eqs. (\ref{nlspec}) and (\ref{nsp})]
due to spectator and nonspectator contributions, respectively,
$\Gamma_{\rm SL}$ [see Eq. (\ref{sl})] and $\Gamma_{\rm lep}$ the
semileptonic and pure leptonic decay widths, respectively. In
units of $\Gamma_0=G_F^2 m_c^5/(192\pi^3)$, we obtain $\Gamma_{\rm
NL,spec}=4.84\,\Gamma_0$, $\Gamma_{\rm SL}=1.02\,\Gamma_0$ and
$\Gamma_{\rm lep}(D_s^+\to\tau\bar\nu_\tau+\mu\bar\nu_\mu)=0.169\,
\Gamma_0$ for $m_c=1.65$ GeV, $m_s=125$ MeV,  $c_1(m_c)=1.30$ and
$c_2(m_c)=-0.57$.

If the momentum of the spectator quark in the $D^+$ meson is
neglected, the destructive Pauli interference in $D^+$ decay is
found to be $\Gamma^{\rm int}_-(D^+)=-8.5\,\Gamma_0$, which
largely overcomes the $c$-quark decay rate $\Gamma_{\rm NL,spec}$.
Consequently, $\Gamma_{\rm tot}(D^+)$ becomes negative, which is
of course of no sense. This indicates that it is mandatory to
invoke the spectator quark to suppress the Pauli interference
effect [see Eq. (\ref{bnonspec})]. On the contrary, the spectator
quark's momentum in the charmed meson will enhance the
$W$-exchange or $W$-annihilation contribution. Since the decay
width of $D^+$ involves a large cancellation between two terms, it
is very sensitive to the parameters $m_c$, $f_D$ and $\la p_q\ra$.
For $f_D=190$ MeV and $\la p_q\ra=350$ MeV, we found that the pole
mass $m_c$ is preferred to be a bit larger. We shall use
$m_c=1.65$ GeV for calculation.

We next proceed to compute the non-spectator effects using Eqs.
(\ref{bnonspec}) and (\ref{para}) and obtain
\be
 \Gamma^{\rm exc}(D^0) &=& (0.46\pm 0.30){\rm \Gamma_0}\,,\nonumber \\
 \Gamma^{\rm int}_-(D^+) &=& -(3.29\pm 0.40){\rm \Gamma_0}\,,\nonumber\\
 \Gamma^{\rm ann}(D^+_s) &=&(0.19\pm 0.13){\rm \Gamma_0}\,,\nonumber \\
 \Gamma^{\rm int}_-(D^+_s) &=& -(0.35\pm 0.05){\rm\Gamma_0}\,,
\en
where the errors come from the uncertainty of $\varepsilon_1$, and
use has been made of $f_{D_s}=240$ MeV. Collecting all the
contributions, we find
\be
\tau(D^0)=0.38\,ps, \qquad \tau(D^+)=0.96\,ps, \qquad
\tau(D_s^+)=0.41\,ps.  \label{tau}
\en
It is clear from our calculations that the lifetime of $D_s^+$ is
longer than that of $D^0$ because the Cabibbo-allowed nonleptonic
annihilation and leptonic contributions to $\Gamma(D_s^+)$ are
compensated by the Cabibbo-suppressed Pauli interference. We also
see that the predicted absolute charmed meson lifetimes are in
general too small compared to experiments \cite{PDG}:
\be
\tau(D^0) = (0.415\pm 0.004)\,ps, \qquad\quad \tau(D^+)=(1.057\pm
0.015)\,ps,
\en
and
\be
\tau(D_s^+)=\cases{ (0.518\pm 0.014\pm 0.007)\, ps &\qquad{\rm
(E791)\cite{E791}}, \cr (0.4863\pm 0.015\pm 0.005)\,ps &\qquad{\rm
(CLEO)\cite{CLEO}}. }
\en
By contrast, the calculated lifetimes of $B$ and $\Lambda_b$
hadrons based on heavy quark expansion are too large compared to
the data (see e.g. \cite{Cheng}).

The charm lifetime ratios followed from Eq. (\ref{tau}) are
\begin{eqnarray}\label{ratios}
&& \frac{\tau(D^+)}{\tau(D^0)}\simeq 2.56\pm 0.52\,, \non \\ &&
\frac{\tau(D^+_s)}{\tau(D^0)}\simeq 1.08\pm 0.04\,.
\end{eqnarray}
Although the lifetime ratio $\tau(D^+)/\tau(D^0)$ is in accordance
with experiment, the predicted ratio for $\tau(D_s^+)/\tau(D^0)$,
which is insensitive to the value of $m_c$, is larger than
previous theoretical estimates \cite{Bigi1,Bigi2} but still
smaller than recent measurements. Nevertheless, this lifetime
ratio could get enhanced if non-vacuum intermediate states
contribute sizably to the four quark condenstate so that $\delta
B$ is nonzero. It is worth remarking that if the nonzero momentum
of the spectator quark is neglected, then the ratio
$\tau(D_s^+)/\tau(D^0)$ will be enhanced to 1.11\,. However, as
stressed in passing, it is meaningless to have a negative lifetime
for the $D^+$.

\acknowledgments This work was supported in part by the National
Science Council of R.O.C. under Grant No. NSC88-2112-M-001-006.


\end{document}